\documentclass[journal,twoside]{IEEEtran}
\usepackage{cite}
\ifCLASSINFOpdf
\else
   \usepackage[dvips]{graphicx}
\fi
\usepackage[cmex10]{amsmath}
\usepackage{amssymb,amsthm}
\usepackage{balance}

\newtheorem{definition}{Definition} 
\newtheorem{theorem}{Theorem}

\begin{document}
\title{Iterative Multiuser Detection and Decoding\\with Spatially Coupled Interleaving} 

\author{Keigo~Takeuchi,~\IEEEmembership{Member,~IEEE} and Shuhei~Horio 
\thanks{Manuscript received June 25, 2013.} 
%The associate editor coordinating the review of this letter 
%and approving it for publication was ***.
\thanks{K.~Takeuchi and S.~Horio are with the Department of Communication 
Engineering and Informatics, the University of Electro-Communications, 
Tokyo 182-8585, Japan (e-mails:ktakeuchi@uec.ac.jp, h1231079@edu.cc.uec.ac.jp).}
\thanks{Digital Object Identifier ***}
}


\markboth{IEEE Wireless communications letters,~Vol.~, No.~, 2013}%
{Takeuchi and Horio:Iterative Multiuser Detection and Decoding with Spatially Coupled Interleaving}

\IEEEpubid{0000--0000/00\$00.00~\copyright~2013 IEEE}


\maketitle

\begin{abstract}
Spatially coupled (SC) interleaving is proposed to improve the performance of 
iterative multiuser detection and decoding (MUDD) for quasi-static fading 
multiple-input multiple-output systems. The linear minimum mean-squared 
error (LMMSE) demodulator is used to reduce the complexity and to avoid 
error propagation. Furthermore, sliding window MUDD is proposed to circumvent 
an increase of the decoding latency due to SC interleaving. 
Theoretical and numerical analyses show that SC interleaving can 
improve the performance of the iterative LMMSE MUDD for regular 
low-density parity-check codes.  
\end{abstract}


%\ifCLASSOPTIONpeerreview
% \fi
%\IEEEpeerreviewmaketitle

\section{Introduction}
\IEEEPARstart{S}{patial} coupling is a sophisticated technique for boosting 
the belief-propagation (BP) decoding threshold up to the optimal 
one~\cite{Kudekar11}. The basic idea of spatially coupled (SC) low-density 
parity-check (LDPC) codes in \cite{Kudekar11} is as follows: An SC LDPC code 
is constructed as a chain of $L$ conventional LDPC codes. The point is to 
introduce an irregular structure at both ends that allows the decoder to 
attain reliable information at the ends. When the code length $M$ of each 
section in the chain is sufficiently long, the reliable information can 
propagate toward the center of the chain regardless of the chain length $L$. 
The influence of the irregularity at both ends is negligible when 
$1\ll L$ ($\ll M$). Consequently, the BP-based iterative decoding can achieve 
the optimal performance. 

In this letter, spatial coupling is utilized to improve the performance of 
BP-based iterative multiuser detection and decoding (MUDD) for quasi-static 
fading multiple-input multiple-output (MIMO) systems. One should not confuse 
the terminology ``spatial coupling'' with coupling in the physical space. 
Coupling is actually made in the time domain.  
Iterative MUDD algorithms based on approximate BP have been proposed 
in ~\cite{Wang99,Boutros02} and analyzed via density evolution in the 
large-system limit where the numbers of transmit and receive antennas tend to 
infinity at the same rate\cite{Caire04}. These low-complexity algorithms can 
achieve excellent decoding performance in the large-system limit compared to 
non-iterative receivers. We propose SC interleaving to improve the performance 
of iterative MUDD via an improvement of the decoding threshold.   

%However, the performance for finite sized systems is far inferior to the 
%asymptotic prediction. That is because infinite diversity can be attained in 
%the large-system limit, whereas only finite diversity is available for the 
%finite sized systems. It was observed in \cite{Takeuchi122} that spatial 
%coupling increases the system size effectively, as well as the improvement of 
%the decoding threshold. 

The main contribution of this letter is to incorporate spatial coupling 
into bit-interleaved coded modulation (BICM), instead of encoding. 
It is possible to combine SC interleaving with any code that 
allows the decoder to use efficient BP decoding. This compatibility of SC 
interleaving is suitable for practical systems that use several codes as 
options. 

%An important example of such codes is SC LDPC codes. 
%One may be concerned about a comparison 
%between the SC interleaver and SC LDPC codes followed by random interleaving. 
%However, making such a comparison fairly is difficult and beyond the scope of 
%this letter.   

As related works, it was proposed in 
\cite{Takeuchi111,Takeuchi122,Schlegel13,Truhachev12} to combine spatial 
coupling with spread spectrum modulation. 
The main difference between this letter and the previous works is that 
practical codes with low decoding complexity are used in this letter, whereas 
the uncoded case~\cite{Takeuchi111,Takeuchi122,Schlegel13} or 
the information-theoretically optimal codes~\cite{Truhachev12} 
were considered in the previous works. In \cite{Truhachev12}, the code rate was 
controlled to avoid the occurrence of error propagation. In this letter, 
we consider LDPC codes with a fixed rate, and investigate the influence of 
error propagation in iterative MUDD.  

\begin{figure}[t]
\begin{center}
\includegraphics[width=\hsize]{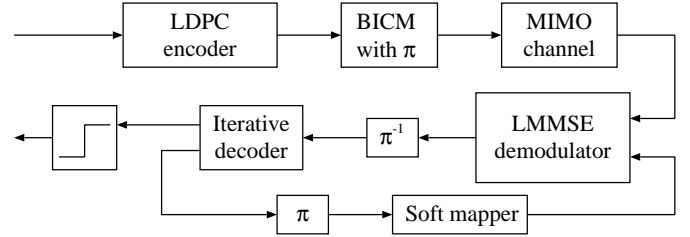}
\end{center}
\caption{
MIMO system. $\pi$ and $\pi^{-1}$ represent the SC interleaver and 
its deinterleaver, respectively.  
}
\label{fig1}
\end{figure}

\IEEEpubidadjcol

\section{System Model}
\subsection{MIMO System} 
We consider an MIMO system with $K$ transmit antennas and 
$N$ receive antennas operating over a frequency-flat quasi-static fading 
channel, shown in Fig.~\ref{fig1}. A binary information stream is encoded with 
a $(d_{\mathrm{v}},d_{\mathrm{c}})$-regular LDPC code of 
code length~$M$~\cite{Richardson08}. After SC interleaving, 
which will be presented shortly, the interleaved stream is modulated and 
divided into $K$ data streams. Gray-mapped quadrature phase shift keying (QPSK) 
$\mathcal{C}=\{a+\mathrm{j}b: a,b=\pm1/\sqrt{2}\}$ is used. 
The obtained data symbols $\{\boldsymbol{x}_{t}=(x_{1,t},\ldots,
x_{K,t})^{\mathrm{T}}\in\mathcal{C}^{K}\}$ are directly transmitted from $K$ 
transmit antennas  at time $t$. The corresponding received 
vector $\boldsymbol{y}_{t}\in\mathbb{C}^{N}$ is given by
\begin{equation} \label{MIMO}
\boldsymbol{y}_{t} = \boldsymbol{H}\boldsymbol{x}_{t} 
+ \boldsymbol{n}_{t}, 
\quad \boldsymbol{n}_{t}\sim\mathcal{CN}(\boldsymbol{0},N_{0}\boldsymbol{I}_{N}). 
\end{equation}
In (\ref{MIMO}), $\{\boldsymbol{n}_{t}\}$ denote independent additive 
white Gaussian noise (AWGN) vectors with covariance $N_{0}\boldsymbol{I}_{N}$.   
The channel matrix $\boldsymbol{H}=(\boldsymbol{h}_{1},\ldots,
\boldsymbol{h}_{K})\in\mathbb{C}^{N\times K}$ is assumed to be independent of the 
time index~$t$ and to be known to the receiver. The former assumption, i.e.\ 
the assumption of quasi-static fading is consistent with the 
latter assumption: It is possible for the receiver to estimate the channel 
matrix by utilizing the known training symbols sent from the transmitter. 
For simplicity, independent and identically distributed (i.i.d.) Rayleigh 
fading is assumed: The channel matrix $\boldsymbol{H}$ has independent 
circularly symmetric complex Gaussian (CSCG) random elements with 
variance $1/K$. Note that there may be dependencies between the elements of 
$\boldsymbol{H}$ in practice. 

\subsection{Spatially Coupled Interleaver} 
\begin{figure}[t]
\begin{center}
\includegraphics[width=\hsize]{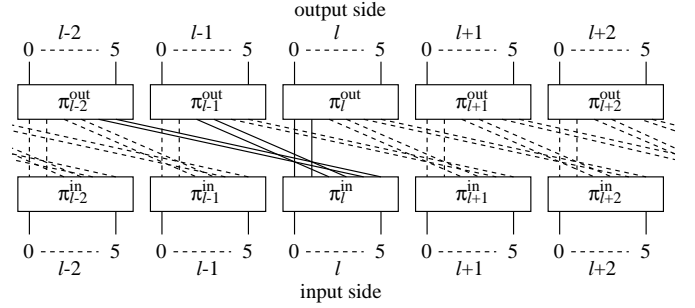}
\end{center}
\caption{
SC interleaver for $M=6$ and $W=3$. $\pi_{l}^{\mathrm{in}}$ and 
$\pi_{l}^{\mathrm{out}}$ denote a random interleaver and a random constrained 
interleaver for section~$l$, respectively. 
}
\label{fig2}
\end{figure}

We shall define an SC interleaver with section size~$M$, 
chain length~$L$, and coupling width~$W$. The section size~$M$ is equal to 
the code length of the used code. The overall length of interleaving is  
$ML$. The chain is coupled circularly, and each section is connected to 
$(W-1)$ neighboring sections uniformly and randomly. We impose a constraint 
under which each data symbol consists of bits in the same codeword, by defining 
a constrained interleaver of length~$M$ that maps consecutive 
integers~$\{2i,2i+1\}$ to consecutive integers~$\{2j,2j+1\}$ for 
$i,j\in\{0,\ldots,M/2-1\}$.  
This constraint simplifies density evolution analysis. 
Input (respectively output) index~$m\in\mathcal{M}=\{0,\ldots,M-1\}$ within 
section~$l\in\mathcal{L}=\{0,\ldots,L-1\}$ corresponds to the 
$(Ml+m)$th input (respectively output) for the SC interleaver. 
See Fig.~\ref{fig2} for an example of the SC interleaver. In particular, 
the SC interleaver with $W=1$ reduces to $L$ uncoupled random interleavers. 
\begin{definition}[SC Interleaver] 
An SC interleaver $\pi$ is a bijection from $\mathcal{M}\times\mathcal{L}$ 
onto $\mathcal{M}\times\mathcal{L}$. Let $\{\pi_{l}^{\mathrm{in}}:
l\in\mathcal{L}\}$ and $\{\pi_{l}^{\mathrm{out}}:l\in\mathcal{L}\}$ denote 
$L$ independent random interleavers of length~$M$ and $L$ 
independent random constrained interleavers of length~$M$, respectively. 
For $(m,l)\in\mathcal{M}\times\mathcal{L}$, 
the SC interleaver $\pi(m,l)$ is given by  
\begin{equation} \label{SC_interleaver} 
\pi(m,l) = (\pi_{l'}^{\mathrm{out}}(\pi_{l}^{\mathrm{in}}(m)),l'), 
\end{equation}
with $l'=(l-(\lfloor \pi_{l}^{\mathrm{in}}(m)/2 \rfloor)_{W})_{L}$, 
in which $(i)_{n}\in\{0,\ldots,n-1\}$ denotes the remainder for the 
division of $i\in\mathbb{Z}$ by $n\in\mathbb{N}$.   
%The SC interleaver is constructed as follows: For each 
%section~$l\in\mathcal{L}$, we pick up a partition 
%$\{\mathcal{M}_{l,w}\subset\mathcal{M}:
%w\in\mathcal{W}=\{0,\ldots,W-1\}\}$ of the index set $\mathcal{M}$ with the 
%same size $|\mathcal{M}_{l,w}|=M/W$ uniformly and randomly. 
%Each subset $\mathcal{M}_{l,w}$ is 
%further decomposed into disjoint subsets with size~$\log_{2}|\mathcal{C}|=2$. 
%The two indices in each decomposed subset are uniformly and randomly mapped to 
%consecutive two indices in the $(l-w)$th section at the output side, which 
%start from some even index. If $(l-w)$ is negative, they are mapped to 
%indices in section~$(l-w+L)$ in the same manner. The procedure above is 
%repeated for all subsets and all sections~$l\in\mathcal{L}$. 
\end{definition}
A known sequence of length $(W-1)M$ is sent in the first $(W-1)$ 
sections. In general, the decoder can decode the codewords in sections close to 
the first sections with smaller error probability than in distant sections. 
When $M$ is sufficiently large, the reliable information at the first 
sections may spreads over the whole system. Consequently, it is possible to 
decode the codewords in distant sections with almost the same error 
probability as in sections close to the first sections. 
 
%Although this causes a rate loss, the known sequence can also be utilized to 
%estimate the channel matrix in practice.  

\section{Iterative MUDD}
\subsection{Sliding Window MUDD} 
SC LDPC codes can be efficiently decoded by sliding window (SW) 
decoding~\cite{Pusane08}. We propose SW MUDD to reduce the decoding delay 
compared to iterative MUDD with parallel scheduling, in which messages are 
collectively sent to the decoder (respectively demodulator) after estimating 
the data symbols (respectively codewords) for {\em all} sections. 
In SW MUDD, the codeword at section~$l$ is 
decoded in the order\footnote{
In order to avoid unnecessary latency, the codewords should be sent in the 
same order after transmission of a known sequence. 
} $l=W-1,L-1,W,L-2,\ldots$. In the decoding stage 
for section~$l$, we update log likelihood ratios (LLRs) exchanged through the 
edges that are connected to the coded bits in section~$l$, shown by the 
solid edges in Fig.~\ref{fig2}, whereas the other LLRs are fixed to the current 
values. In each iteration, the demodulator calculates LLRs and sends them to 
the decoder, which uses the passed LLRs to calculate extrinsic LLRs to 
be fed back to the demodulator. After convergence or $I$ iterations, 
the decoder outputs the decoding results, and the SW MUDD proceeds to the 
next stage. Note that the SW MUDD is suboptimal, since we do not update LLRs 
passed through the edges that are connected to the following sections. 

The extrinsic LLRs passed from the decoder toward the demodulator are 
calculated by using the conventional sum-product algorithm with 
the number of iterations~$J$~\cite{Richardson08}. The LLRs passed in the 
opposite direction are updated by using the linear minimum 
mean-squared error (LMMSE) demodulator, 
since the optimal demodulator is infeasible in terms of the complexity.  
In this letter, we refer to iterations in the decoder and in the MUDD 
as inner and outer iterations, respectively. 

\subsection{LMMSE Demodulator}
Let $L_{k,t}^{\mathrm{dec}}\in\mathbb{C}$ denote the a priori {\em complex} LLR for 
the data symbol $x_{k,t}\in\mathcal{C}$ fed back from the decoder 
in an outer iteration round. 
The real and imaginary parts of $L_{k,t}^{\mathrm{dec}}$ correspond to the LLRs 
for those of the data symbol, respectively. 
In the initial outer iteration for the SW scheduling, the a priori LLR is 
equal to that passed from the decoder in the preceding stage if it exists. 
Otherwise, the LLR is set to zero. 
The corresponding a priori probability $p(x_{k,t})$ is given by 
$p(x_{k,t})=p(\Re[x_{k,t}])p(\Im[x_{k,t}])$, in which 
\begin{equation} \label{prior} 
p\left(
 \Re[x_{k,t}]=\pm \frac{1}{\sqrt{2}}
\right) 
= \frac{\mathrm{e}^{\pm\Re[L_{k,t}^{\mathrm{dec}}]/2}}
{\mathrm{e}^{\Re[L_{k,t}^{\mathrm{dec}}]/2} + \mathrm{e}^{-\Re[L_{k,t}^{\mathrm{dec}}]/2}}, 
\end{equation}
where a positive LLR implies that $\Re[x_{k,t}]=1/\sqrt{2}$ is more likely.  
The real part of the mean $\hat{x}_{k,t}$ with respect to $p(x_{k,t})$ is 
given by $\Re[\hat{x}_{k,t}]=2^{-1/2}\tanh(\Re[L_{k,t}^{\mathrm{dec}}]/2)$. 
The a priori probability and mean of $\Im[x_{k,t}]$ are defined 
in the same manner. 
Thus, the a priori variance $(1-|\hat{x}_{k,t}|^{2})$ of the QPSK symbol 
$x_{k,t}$ is given by 
$\sigma^{2}(\Re[L_{k,t}^{\mathrm{dec}}],\Im[L_{k,t}^{\mathrm{dec}}])$, with 
\begin{equation} \label{prior_variance} 
\sigma^{2}(L_{\mathrm{r}},L_{\mathrm{i}}) 
= 1 - \frac{1}{2}\left\{
 \tanh^{2}\left(
  \frac{L_{\mathrm{r}}}{2}
 \right) 
 + \tanh^{2}\left(
  \frac{L_{\mathrm{i}}}{2}
 \right)
\right\}. 
\end{equation} 

We focus on the $k$th symbol at time~$t$, and shall derive the LMMSE 
estimation of $x_{k,t}$ based on the 
{\em extrinsic} information $\{p(x_{k',t}):k'\neq k\}$. 
The use of the true a priori probability $p(x_{k',t})$ results in the 
optimal nonlinear demodulator with high complexity. In order to reduce the 
complexity, the a priori probabilities $p(x_{k',t})$ for all $k'\neq k$ 
are approximated by proper complex Gaussian distributions with mean 
$\hat{x}_{k',t}$ and variance $(1-|\hat{x}_{k',t}|^{2})$ that are equal 
to those of $x_{k',t}$ for $p(x_{k',t})$, respectively. On the other hand,  
$p(x_{k,t})$ is approximated by a CSCG distribution with unit variance. 
These approximations result in the approximate posterior probability density 
function (pdf) of $x_{k,t}$ given by 
\begin{equation} \label{posterior} 
p(x_{k,t}| \boldsymbol{y}_{t},\boldsymbol{H}) 
= \frac{1}{\pi\xi_{k,t}}\exp\left(
 - \frac{|x_{k,t} - \bar{x}_{k,t}|^{2}}{\xi_{k,t}} 
\right), 
\end{equation}
with
\begin{equation} \label{estimator} 
\bar{x}_{k,t} 
= \xi_{k,t}\boldsymbol{h}_{k}^{\mathrm{H}}
\boldsymbol{\Sigma}_{k,t}^{-1}\left(
 \boldsymbol{y}_{t} - \sum_{k'\neq k}\boldsymbol{h}_{k'}\hat{x}_{k',t}
\right), 
\end{equation}
\begin{equation}
\xi_{k,t} 
= \left(
 1 + \boldsymbol{h}_{k}^{\mathrm{H}}\boldsymbol{\Sigma}_{k,t}^{-1}
 \boldsymbol{h}_{k}
\right)^{-1}. 
\end{equation}
In these expressions, $\boldsymbol{\Sigma}_{k,t}$ is given by 
\begin{equation} \label{Sigma} 
\boldsymbol{\Sigma}_{k,t}  
= N_{0}\boldsymbol{I}_{N} 
+ \sum_{k'\neq k}(1-|\hat{x}_{k',t}|^{2})\boldsymbol{h}_{k'}
\boldsymbol{h}_{k'}^{\mathrm{H}}. 
\end{equation}
Expression~(\ref{posterior}) implies that the complex LLR 
$L_{k,t}^{\mathrm{dem}}\in\mathbb{C}$ for $x_{k,t}$ sent from the LMMSE demodulator 
to the decoder is given by 
\begin{equation} \label{LLR_linear} 
L_{k,t}^{\mathrm{dem}}
= 2\sqrt{2}\boldsymbol{h}_{k}^{\mathrm{H}}
\boldsymbol{\Sigma}_{k,t}^{-1}\left(
 \boldsymbol{y}_{t} - \sum_{k'\neq k}\boldsymbol{h}_{k'}\hat{x}_{k',t}
\right), 
\end{equation}
of which real and imaginary parts correspond to the LLRs for those 
of $x_{k,t}$, respectively. 

Unless $N_{0}$ is zero, {\em soft} information about the data symbols, 
i.e.\ a finite LLR~(\ref{LLR_linear}) is fed forward to the decoder. 
Note that the LMMSE demodulator reduces to the zero-forcing (ZF) demodulator 
when $N_{0}$ in (\ref{Sigma}) is approximated by a sufficiently small value. 
Since the second term in (\ref{Sigma}) is not invertible for $N>K-1$,   
the LLR~(\ref{LLR_linear}) diverges when the ZF demodulator is 
used. In other words, {\em hard} information about the data symbols is sent 
to the decoder. The hard information may result in error propagation, so that 
the LMMSE demodulator is used to avoid error propagation in this letter. 

\section{Density Evolution}
\subsection{Asymptotic Analysis}
We follow \cite{Caire04} to present the density evolution of the iterative 
MUDD in the large-system limit after taking the infinite code length 
limit~$M\to\infty$. In the large-system limit, the numbers of transmit and 
receive antennas tend to infinity with their ratio $\alpha=K/N$ kept constant. 
It is known that the large-system analysis can provide a good prediction 
for the starting location of the so-called {\em waterfall} regime.  

Let $p_{l}^{\mathrm{dec}}(L)$ denote the asymptotic pdf of the real LLRs emitted 
from the decoder for section~$l$ as $M\to\infty$. 
%In the decoding stage~$l$ for the SW scheduling, the 
%pdfs $p_{l'}^{\mathrm{dec}}(L)$ for the preceding sections~$l'<l$ are 
%fixed to $p_{l'}^{\mathrm{dec}}(L)=p_{l'}^{\mathrm{dec}}(\cdot;l',I)$ at the 
%corresponding stages, whereas $p_{l'}^{\mathrm{dec}}(L;l,i)$ for the following 
%stages~$l'$ are set to the Dirac delta function $\delta(L)$. 
The pdf $p_{l}^{\mathrm{dec}}(L)$ can be analyzed with the Gaussian approximation 
of the LLRs~\cite{Richardson08}. 
Thus, we mainly present the large-system analysis of the demodulator.  

\subsection{LMMSE Demodulator} \label{sec4-B}
The analysis of the LMMSE demodulator is based on \cite{Tse99}. 
We focus on section~$l$, and suppose that $\{p_{l'}^{\mathrm{dec}}(L):l'=l,\ldots,
l+W-1\}$ have been fed back from the decoder.  
Let $(k,t)$ denote any couple that corresponds to indices included in 
section~$l$ at the output side of the SC interleaver. It is proved that the 
LLR~(\ref{LLR_linear}) sent to the decoder is statistically equivalent to 
that for the interference-free complex AWGN channel 
\begin{equation} \label{AWGN} 
z_{l} = x_{l} + n_{l}, 
\quad n_{l}\sim\mathcal{CN}(0,\sigma_{l}^{2}), 
\end{equation}
with $x_{l}\in\mathcal{C}$ denoting the input symbol for section~$l$. 
In (\ref{AWGN}), the noise variance~$\sigma_{l}^{2}$ will be defined 
shortly. The complex LLR $L_{l}^{\mathrm{AWGN}}\in\mathbb{C}$ for the AWGN 
channel~(\ref{AWGN}) with the uniform a priori probability 
$p(x_{l})=1/|\mathcal{C}|$ is given by 
\begin{equation} \label{LLR_AWGN} 
L_{l}^{\mathrm{AWGN}} = 2\sqrt{2}\frac{z_{l}}{\sigma_{l}^{2}}. 
\end{equation}
The distribution of the LLR~(\ref{LLR_AWGN}) is statistically equivalent to 
that of the original LLR~(\ref{LLR_linear}) in the large-system limit, 
when $\sigma_{l}^{2}$ is given as the solution to a fixed-point equation. 
\begin{theorem} \label{theorem1} 
%For iteration~$i$ in decoding stage~$l$ of the SW MUDD. 
Focus on section~$l$, and suppose that $\{p_{l'}^{\mathrm{dec}}(L):l'=l,\ldots,
l+W-1\}$ have been fed back from the decoder.  
For any couple $(k,t)$ that corresponds to indices included in 
section~$l$, the LLR~(\ref{LLR_linear}) given $\boldsymbol{H}$, 
$\{\hat{x}_{k',t}\}$, and $x_{k,t}=x$ converges in 
distribution to (\ref{LLR_AWGN}) given $x_{l}=x$ with probability~$1$ 
in the large-system limit after taking $M\to\infty$. 
In evaluating (\ref{LLR_AWGN}), $\sigma_{l}^{2}$ is 
given by the solution to the fixed-point equation,
\begin{equation} \label{fixed_point}
\sigma_{l}^{2}
= \alpha\left(
 N_{0} + \frac{1}{W}\sum_{w=0}^{W-1}\mathrm{MSE}_{l+w}(\sigma^{2}_{l}) 
\right), 
\end{equation}
with 
\begin{equation} \label{MSE} 
\mathrm{MSE}_{l'}(\sigma_{l}^{2}) 
= \int_{\mathbb{R}^{2}} 
\frac{\sigma^{2}(L_{\mathrm{r}},L_{\mathrm{i}})\sigma_{l}^{2}}
{\sigma^{2}(L_{\mathrm{r}},L_{\mathrm{i}}) + \sigma_{l}^{2}}
p_{l'}^{\mathrm{dec}}(L_{\mathrm{r}})p_{l'}^{\mathrm{dec}}(L_{\mathrm{i}})
dL_{\mathrm{r}}dL_{\mathrm{i}}, 
\end{equation}
where the a priori variance $\sigma^{2}(L_{\mathrm{r}},L_{\mathrm{i}})$ 
is given by (\ref{prior_variance}).  
\end{theorem} 
\begin{IEEEproof}
See Appendix~\ref{proof_theorem1}.
\end{IEEEproof}
The function~(\ref{MSE}) corresponds to the average power of the interference 
due to the data symbols associated with the $l'$th codeword. Thus, the 
interference power tends to zero when the mass of $p_{l'}^{\mathrm{dec}}(L)$ 
concentrates at $\pm\infty$. This situation occurs when the $l'$th decoder 
sends the correct hard decision.  

Theorem~\ref{theorem1} implies that the asymptotic multiuser efficiency (ME) 
for section~$l$ is given by $\alpha N_{0}/\sigma_{l}^{2}$, and that the analysis 
of the decoder for section~$l$ reduces to that of decoder for the 
LDPC-coded AWGN channel~(\ref{AWGN}) with $W$ signal-to-noise ratio (SNR) 
levels $\{1/\sigma_{l'}^{2}:l'=l-(W-1),\ldots,l\}$. This problem can be 
solved with the Gaussian approximation of the LLRs~\cite{Richardson08}. 
See Appendix~\ref{decoder_DE} for the details. 

\section{Numerical Results and Concluding Remarks}
The performance of the SC interleaving is compared to that of 
the conventional random interleaving with $W=1$. In all numerical results, 
we used $(3,6)$-regular LDPC codes~\cite{Richardson08}. 

\begin{figure}[t]
\begin{center}
\includegraphics[width=\hsize]{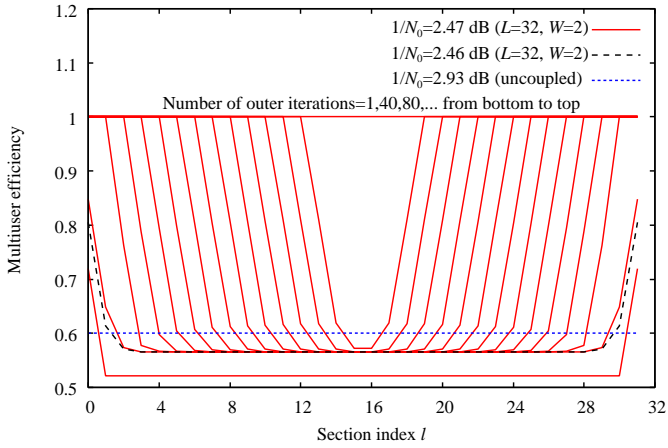}
\end{center}
\caption{
Multiuser efficiency for the iterative LMMSE MUDD 
with parallel scheduling for $\alpha=K/N=1$ and the number of inner 
iterations~$J=100$.  
}
\label{fig3}
\end{figure}

Figure~\ref{fig3} shows the evolution of the asymptotic ME based on 
Theorem~\ref{theorem1}. We used the parallel scheduling in order to clarify 
the behavior of the iterative MUDD. 
ME close to one implies that the inter-stream interference has been 
eliminated. Consequently, the systems can enjoy the interference-free 
performance. The ME for the conventional interleaving tends to 
a value distant from one after sufficiently many iterations 
when SNR $1/N_{0}=2.93$~dB. On the other hand, the ME for the SC interleaving 
can still converge to one for $1/N_{0}=2.47$~dB, whereas it cannot for 
$1/N_{0}=2.46$~dB. These observations imply that the decoding threshold 
is between $2.46$~dB and $2.47$~dB, which is defined as the minimum SNR 
such that the ME converges to one after infinite outer iterations.  

\begin{table}[t]
\begin{center}
\caption{
Decoding thresholds for sufficiently large~$L$, $\alpha=K/N=1$, and 
the number of inner iterations~$J=100$. 
}
\label{table1}
\begin{tabular}{|c||c|c|c|c|}
\hline
& $W=1$ & $W=2$ & $W=3$ & $W=4$ \\ 
\hline\hline
Parallel & $2.94$~dB & $2.47$~dB & $2.36$~dB & $2.31$~dB \\ 
\hline
SW & $2.94$~dB & $2.60$~dB & $2.55$~dB & $2.53$~dB \\ 
\hline
\end{tabular}
\end{center}
\end{table}

Table~\ref{table1} lists the decoding thresholds for the parallel and 
SW scheduling. The chain length~$L$ was set to sufficiently large values to 
eliminate the influence of the rate loss due to spatial coupling. 
We find that the SC interleaving with $W\geq2$ 
can improve the decoding threshold compared to the conventional interleaving 
with $W=1$. Furthermore, the SW scheduling with a decoding delay of $O(W)$ is 
slightly inferior to the parallel scheduling with a delay of $O(L)$. 
This implies that there is a tradeoff between the performance and the delay. 

\begin{figure}[t]
\begin{center}
\includegraphics[width=\hsize]{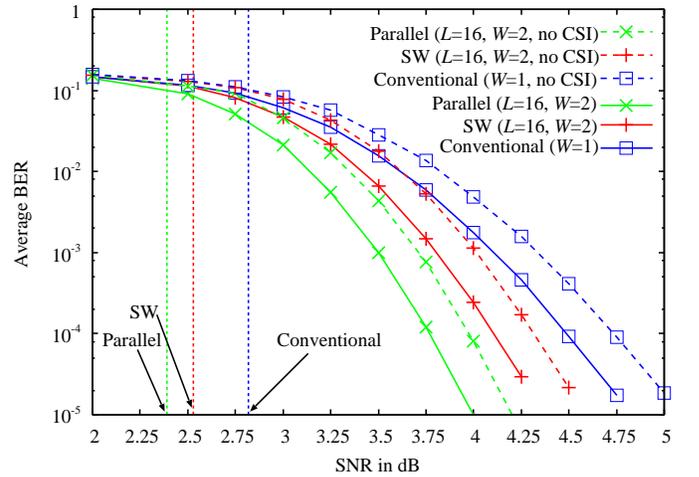}
\end{center}
\caption{
BER versus SNR~$1/N_{0}$ for the $16\times16$ MIMO system. Code 
length~$M=2048$, and the numbers of inner and outer iterations~$J\leq100$ 
and $I\leq300$, respectively. The vertical lines show the decoding 
thresholds for the three schemes.   
}
\label{fig4}
\end{figure}

Figure~\ref{fig4} shows the average bit error rates (BERs) in decoding 
for the $16\times 16$ MIMO system. Note that the iterative MUDD converges 
quickly for the SNR regime above the decoding thresholds, although the 
maximum numbers of inner and outer iterations were set to $100$ and 
$300$, respectively. As a fair comparison between the conventional and SC 
interleavers in terms of the overall rate, 
we also plotted the BERs for the case of no channel state information (CSI): 
A random training binary sequence of length~$16384$ was transmitted for the 
(non-iterative) LMMSE channel estimation in the MIMO system with the 
conventional interleaving. For the SC interleaving, on the other hand, 
$12.5$~\% of the sequence was used as the training sequence for spatial 
coupling, and the remaining sequence was utilized for the LMMSE channel 
estimation. 
We find that the SC interleaving can provide performance gains of $0.5$~dB 
and $0.8$~dB for the SW and parallel scheduling, respectively, compared to 
the conventional random interleaving.

%\section*{Acknowledgment}

%\appendices
\appendices
\section{Proof of Theorem~\ref{theorem1}}
\label{proof_theorem1} 
In the proof of Theorem~\ref{theorem1}, we treat the channel matrix 
$\boldsymbol{H}$ and the soft decisions $\{\hat{x}_{k',t}\}$ as deterministic 
variables, and omit conditioning with respect to these variables.  
It is known that the LLR for linear receivers converges in 
distribution to a Gaussian random variable with probability~$1$ 
in the large-system limit, e.g., see \cite{Guo02,Guo05}. 
Thus, it is sufficient to evaluate the conditional mean 
and variance of the LLR~(\ref{LLR_linear}). 

We first calculate the conditional mean 
$\mathbb{E}[L_{k,t}^{\mathrm{dem}}|x_{k,t}=x]$. 
Substituting (\ref{MIMO}) into (\ref{LLR_linear}) yields   
\begin{equation} \label{LLR_tmp1} 
\frac{L_{k,t}^{\mathrm{dem}}}{2\sqrt{2}}  
= \boldsymbol{c}_{k,t}^{\mathrm{H}}\boldsymbol{h}_{k}x_{k,t} 
+ \sum_{k'\neq k}\boldsymbol{c}_{k,t}^{\mathrm{H}}
\boldsymbol{h}_{k'}(x_{k',t}-\hat{x}_{k',t}) 
+ \boldsymbol{c}_{k,t}^{\mathrm{H}}\boldsymbol{n}_{t}, 
\end{equation}
with $\boldsymbol{c}_{k,t}=\boldsymbol{\Sigma}_{k,t}^{-1}\boldsymbol{h}_{k}$ 
denoting the LMMSE filter. Expression~(\ref{LLR_tmp1}) implies 
\begin{equation}
\mathbb{E}[L_{k,t}^{\mathrm{dem}}|x_{k,t}=x]=2\sqrt{2}
\boldsymbol{h}_{k}^{\mathrm{H}}\boldsymbol{\Sigma}_{k,t}^{-1} 
\boldsymbol{h}_{k}x, 
\end{equation}
where we have used the assumption that the a priori mean of $x_{k',t}$ is equal 
to $\hat{x}_{k',t}$ for all $k'\neq k$. 

We next evaluate the conditional variance 
$\mathbb{V}[L_{k,t}^{\mathrm{dem}}|x_{k,t}]$. Using the fact that 
$\{x_{k',t}\}$ are regarded as independent random variables for all $k'\neq k$ 
in the limit $M\to\infty$, because of random interleaving, we obtain  
\begin{IEEEeqnarray}{rl}
\mathbb{V}\left[
 \left. 
  \frac{L_{k,t}^{\mathrm{dem}}}{2\sqrt{2}}
 \right|x_{k,t}
\right] 
=& \sum_{k'\neq k}|\boldsymbol{c}_{k,t}^{\mathrm{H}}\boldsymbol{h}_{k'}|^{2}
 (1-|\hat{x}_{k',t}|^{2})
 + N_{0}\|\boldsymbol{c}_{k,t}\|^{2} \nonumber \\ 
=& \boldsymbol{c}_{k,t}^{\mathrm{H}}\boldsymbol{\Sigma}_{k,t}\boldsymbol{c}_{k,t}, 
\label{var} 
\end{IEEEeqnarray} 
with (\ref{Sigma}). Substituting 
$\boldsymbol{c}_{k,t}=\boldsymbol{\Sigma}_{k,t}^{-1}\boldsymbol{h}_{k}$, we 
obtain 
\begin{equation}
\mathbb{V}[L_{k,t}^{\mathrm{dem}}|x_{k,t}] 
= 8\boldsymbol{h}_{k}^{\mathrm{H}}\boldsymbol{\Sigma}_{k,t}^{-1} 
\boldsymbol{h}_{k}. 
\end{equation}

As observed from (\ref{LLR_AWGN}), on the other hand, 
the LLR~(\ref{LLR_AWGN}) conditioned on $x_{l}=x$ for the AWGN channel has  
mean~$2\sqrt{2}x/\sigma_{l}^{2}$ and variance $8/\sigma_{l}^{2}$. Thus, 
it is sufficient to prove that $\boldsymbol{h}_{k}^{\mathrm{H}}
\boldsymbol{\Sigma}_{k,t}^{-1}\boldsymbol{h}_{k}$ converges to $1/\sigma_{l}^{2}$ 
in the large-system limit, given by the solution to the fixed-point 
equation~(\ref{fixed_point}).   

It is worth noting that the quantity $\boldsymbol{h}_{k}^{\mathrm{H}}
\boldsymbol{\Sigma}_{k,t}^{-1}\boldsymbol{h}_{k}$ is equal to 
the signal-to-interference ratio (SIR) $\mathrm{sir}_{k,t}$ 
for the LLR~(\ref{LLR_linear}) in the limit $M\to\infty$. 
In fact, from (\ref{LLR_tmp1}) we obtain   
\begin{equation} \label{sir} 
\mathrm{sir}_{k,t} 
= \frac{8|\boldsymbol{c}_{k,t}^{\mathrm{H}}\boldsymbol{h}_{k}|^{2}}
{\mathbb{V}[L_{k,t}^{\mathrm{dem}}|x_{k,t}]} 
=\boldsymbol{h}_{k}^{\mathrm{H}}\boldsymbol{\Sigma}_{k,t}^{-1}
\boldsymbol{h}_{k}, 
\end{equation}
where we have used $\boldsymbol{c}_{k,t}=
\boldsymbol{\Sigma}_{k,t}^{-1}\boldsymbol{h}_{k}$ and (\ref{var}). 
The SIR~(\ref{sir}) depends on the channel matrix $\boldsymbol{H}$ and 
the soft decisions $\{\hat{x}_{k',t}\}$ via (\ref{Sigma}). 
Tse and Hanly~\cite{Tse99} used random matrix theory to prove\footnote{
More precisely, they considered the case of real random variables. 
However, the result is easily generalized to the case of 
complex random variables.} that the SIR~(\ref{sir}) 
converges {\em in probability} to a deterministic value in the large-system 
limit. Here, we shall present a bit stronger statement~\cite{Tulino04}. 
\begin{theorem}[\cite{Tse99}] \label{theorem2} 
Suppose that $\sigma_{l}^{2}$ is the solution to the fixed-point equation 
\begin{equation} \label{FP} 
\sigma_{l}^{2} = \alpha\left(
 N_{0} + \int\frac{x\sigma_{l}^{2}}{x + \sigma_{l}^{2}}
 dF(x)
\right), 
\end{equation}
where $F(x)$ represents the limiting empirical distribution of the a priori 
variances $\{1-|\hat{x}_{k',t}|^{2}\}$, 
\begin{equation} \label{empirical} 
F(x) 
= \lim_{K\to\infty}\frac{1}{K-1}\sum_{k'\neq k}\chi\left(
 1-|\hat{x}_{k',t}|^{2}\leq x
\right), 
\end{equation}
with $\chi$ denoting the indicator function. Then, 
the SIR~(\ref{sir}) converges almost surely to $1/\sigma_{l}^{2}$ 
in the large-system limit. 
\end{theorem} 
Note that the fixed-point equation~(\ref{FP}) for the MIMO system is slightly 
different from the so-called Tse-Hanly equation~\cite{Tse99} for code-division 
multiple-access (CDMA) systems, because of the difference in power 
normalization. 

In order to complete the proof of Theorem~\ref{theorem1}, we evaluate 
the limiting empirical distribution~(\ref{empirical}). 
From (\ref{prior_variance}), 
(\ref{empirical}) reduces to 
\begin{equation} \label{empirical_tmp1} 
F(x) 
= \lim_{K\to\infty}\frac{1}{K-1}\sum_{k'\neq k}\chi\left(
 \sigma^{2}(\Re[L_{k',t}^{\mathrm{dec}}],\Im[L_{k',t}^{\mathrm{dec}}])\leq x
\right). 
\end{equation}
Recall that we are focusing on section~$l$ at the output side of the SC 
interleaver. From the construction of the SC interleaver, the $W$ decoders 
from sections~$l$ to $l+W-1$ feed the LLRs back to the demodulator 
with equal probability in the large-system limit. Furthermore, the assumption 
of the random bit-interleaving implies that the LLRs 
$\{\Re[L_{k',t}^{\mathrm{dec}}],\Im[L_{k',t}^{\mathrm{dec}}]:k'\neq k\}$ 
are independent random variables, since we have first taken the limit 
$M\to\infty$. From the law of large numbers, the empirical 
distribution~(\ref{empirical_tmp1}) converges almost surely to 
\begin{IEEEeqnarray}{r}
F(x) = \frac{1}{W}\sum_{w=0}^{W-1}
\int_{\mathbb{R}^{2}}\chi\left(
 \sigma^{2}(L_{\mathrm{r}},L_{\mathrm{i}})\leq x
\right) \nonumber \\ 
\cdot p_{l+w}^{\mathrm{dec}}(L_{\mathrm{r}})
p_{l+w}^{\mathrm{dec}}(L_{\mathrm{i}})dL_{\mathrm{r}}dL_{\mathrm{i}}, 
\label{empirical2} 
\end{IEEEeqnarray}
which implies that the fixed-point equation~(\ref{FP}) reduces to 
(\ref{fixed_point}).  
Note that the subscripts of the two pdfs in (\ref{empirical2}) coincide with 
each other, since each data symbol consists of bits in the same codeword.

\section{Density Evolution for Decoder} 
\label{decoder_DE}  
We shall present the DE analysis for the $l$th decoder. 
In Section~\ref{sec4-B}, we have proved that the analysis of the $l$th decoder 
reduces to that of the decoder for the LDPC-coded complex AWGN channel 
with $W$ SNR levels $\{1/\sigma_{l'}^{2}:l'=l-(W-1),\ldots,l\}$. 
Since it is infeasible to trace the exact distribution of LLRs, 
we follow \cite{Richardson08} to approximate the distributions of the LLRs by 
Gaussian distributions. As shown in Appendix~\ref{proof_theorem1}, 
the LLR~(\ref{LLR_AWGN}) conditioned on $x_{l}$ has  
mean~$2\sqrt{2}x_{l}/\sigma_{l}^{2}$ and variance $8/\sigma_{l}^{2}$. 
Since QPSK is used, the products $\Re[L_{l}^{\mathrm{AWGN}}]\Re[x_{l}]$ and 
$\Im[L_{l}^{\mathrm{AWGN}}]\Im[x_{l}]$ are independent of each other, and follow 
the real Gaussian distribution with mean $2/\sigma_{l}^{2}$ and 
variance~$4/\sigma_{l}^{2}$. In the Gaussian approximation of LLRs, 
the distributions of the LLRs are approximated by this constrained Gaussian 
distributions.  

Without loss of generality, we assume transmission of all-zero codeword. 
We follow \cite{Richardson08} to use the entropy~$h=\psi(m)$ as the parameter 
that determines the constrained Gaussian distribution, instead of mean~$m>0$, 
given by
\begin{equation} \label{psi} 
\psi(m) = \int_{\mathbb{R}}S\left(
 \frac{\mathrm{e}^{L/2}}{\mathrm{e}^{L/2}+\mathrm{e}^{-L/2}}
\right)
\frac{1}{\sqrt{4\pi m}}\mathrm{e}^{-\frac{(L-m)^{2}}{4m}}dL, 
\end{equation}
where $S(p)$ denotes the binary entropy function 
\begin{equation}
S(p) = -p\log_{2}p - (1-p)\log_{2}(1-p). 
\end{equation}
As seen from (\ref{prior}), the entropy~$h=\psi(m)$ is regarded as the average 
entropy of a binary random variable characterized by a Gaussian-distributed 
LLR~$L$ with mean~$m$ and variance $2m$. Since (\ref{psi}) is monotonically 
decreasing, the inverse function $m=\psi^{-1}(h)$ exists.  

Let $h_{\mathrm{c},l}^{(j)}$ denote the entropy for the LLR emitted from each check 
node in inner iteration~$j$. Recall that we are focusing on the $l$th decoder. 
We approximate the pdf of the LLR 
by the Gaussian pdf with mean $\psi^{-1}(h_{\mathrm{c},l}^{(j)})$ and 
variance $2\psi^{-1}(h_{\mathrm{c},l}^{(j)})$. Since each variable node sends 
to a check node the sum of LLRs sent by the demodulator and by the 
other $(d_{\mathrm{v}}-1)$ check nodes connected to the variable node,  
we evaluate the entropy $h_{\mathrm{v},l,l'}^{(j)}$ for 
the LLR emitted from a variable node that is connected to the AWGN channel 
with SNR~$1/\sigma_{l'}^{2}$ as 
\begin{equation} \label{each_entropy_v} 
h_{\mathrm{v},l,l'}^{(j)} = 
\psi\left(
 \frac{2}{\sigma_{l'}^{2}} + (d_{\mathrm{v}}-1)\psi^{-1}(h_{\mathrm{c},l}^{(j-1)})
\right),
\end{equation}
with $h_{\mathrm{c},l}^{(0)}=1$. Thus, the average entropy $h_{\mathrm{v},l}^{(j)}$ 
emitted from the variable nodes is given by 
\begin{equation} \label{entropy_v}
h_{\mathrm{v},l}^{(j)} = \frac{1}{W}\sum_{w=0}^{W-1} 
\psi\left(
 \frac{2}{\sigma_{l-w}^{2}} + (d_{\mathrm{v}}-1)\psi^{-1}(h_{\mathrm{c},l}^{(j-1)})
\right). 
\end{equation}
Here, we approximate the distribution of the LLRs emitted from the variable 
nodes by $\mathcal{N}(\psi^{-1}(h_{\mathrm{v},l}^{(j)}),
2\psi^{-1}(h_{\mathrm{v},l}^{(j)}))$, 
although the {\em true} distribution is the mixture of Gaussian distributions 
under the first Gaussian approximation. Note that the same approximation is 
made in the DE analysis of irregular LDPC codes. 

In order to calculate the entropy $h_{\mathrm{c},l}^{(j)}$, 
we use the duality between variable nodes with entropy~$h$ and 
check nodes with entropy~$(1-h)$~\cite{Richardson08}. 
Exchanging the roles of variable nodes and check nodes, and repeating 
the derivation of (\ref{each_entropy_v}), we obtain  
\begin{equation} \label{entropy_c} 
h_{\mathrm{c},l}^{(j)} 
= 1 - \psi\left(
 (d_{\mathrm{c}}-1)\psi^{-1}(1-h_{\mathrm{v},l}^{(j)})
\right). 
\end{equation} 

\balance

The two expressions~(\ref{entropy_v}) and (\ref{entropy_c}) correspond to 
the DE equations for the decoder. The entropy $h_{l}^{\mathrm{dec}}$ for the LLR 
passed from the $l$th decoder to the demodulator is given by 
\begin{equation}
h_{l}^{\mathrm{dec}} = \psi\left(
 d_{\mathrm{v}}\psi^{-1}(h_{\mathrm{c},l}^{(J)}) 
\right), 
\end{equation}
where $J$ denotes the total number of inner iterations. This implies that the 
asymptotic pdf $p_{l}^{\mathrm{dec}}(L)$ emitted from the $l$th decoder is 
given by 
\begin{equation}
p_{l}^{\mathrm{dec}}(L) 
= \frac{1}{\sqrt{4\pi\psi^{-1}(h_{l}^{\mathrm{dec}})}}
\mathrm{e}^{
 - \frac{(L-\psi^{-1}(h_{l}^{\mathrm{dec}}))^{2}}{4\psi^{-1}(h_{l}^{\mathrm{dec}})}
}, 
\end{equation}
where $\psi^{-1}$ denotes the inverse function of (\ref{psi}).

\ifCLASSOPTIONcaptionsoff
  \newpage
\fi



\bibliographystyle{IEEEtran}
\bibliography{IEEEabrv,kt-coml2013}
\end{document}